\documentclass[
aps,
prd,
showpacs,
tightenlines,
nofootinbib,
floatfix]
{revtex4}
\usepackage{graphicx,
longtable
}

\begin{document}
\title{	Observables sensitive to absolute neutrino masses.~II}
\author{G.L.~Fogli$^{1,2}$, E.~Lisi$^2$, A.~Marrone$^{1,2}$, A.~Melchiorri$^3$, 
		A.~Palazzo$^{4}$, A.M.~Rotunno$^{1,2}$, P.~Serra$^5$, J.~Silk$^6$, A.~Slosar$^7$}
\address{
		$^1$~Dipartimento di Fisica, Universit\`a di Bari, Via Amendola 173, 70126, Bari, Italy	\smallskip \\ 
		$^2$~Istituto Nazionale di Fisica Nucleare (INFN), Sezione di Bari,
		Via Orabona 4, 70126 Bari, Italy 										\smallskip \\
		$^3$~Dipartimento di Fisica and Sezione INFN, Universit\`a
		degli Studi di Roma ``La Sapienza'', P.le Aldo Moro 5, 00185, Rome, Italy 	\smallskip \\
		$^4$~AHEP Group, Institut de F\'isica Corpuscular, CSIC/Universitat de Val\`encia,
		Edifici Instituts d'Investigaci\'o, Apt.\ 22085, 46071 Val\`encia, Spain 	\smallskip \\
		$^5$~Department\ of Physics and Astronomy, University of California, 
		Irvine, CA 92697-4575 													\smallskip \\
		$^6$~Astrophysics, Denys Wilkinson Building, Keble Road, 
		Oxford, OX1 3RH, United Kingdom 											\smallskip \\
		$^7$~Berkeley Center for Cosmological Physics, Physics Department, 
		University of California, Berkeley CA 94720								\medskip\medskip}

\begin{abstract}
\medskip
In this followup to Phys.\ Rev.\ D {\bf 75}, 053001 (2007) [arXiv:hep-ph/0608060]
we report updated constraints on neutrino mass-mixing parameters, 
in light of recent neutrino oscillation data (KamLAND, SNO, and MINOS) and    
cosmological observations (WMAP 5-year and other data). We discuss their 
interplay with the final $0\nu2\beta$ decay results in $^{76}$Ge claimed  by part of the
Heidelberg-Moscow Collaboration, using recent evaluations of the corresponding nuclear 
matrix elements, and their uncertainties. We also comment on the
$0\nu2\beta$ limits in $^{130}$Te recently set by Cuoricino, and on prospective limits 
or signals from the KATRIN experiment.
\end{abstract}
 \maketitle


{\em Introduction.}
This paper is meant as a followup to the article \cite{Fogli2006}
where, building upon previous work \cite{Fogli2004},
we presented constraints on the neutrino mass-squared differences
($\delta m^2,\,\Delta m^2$) and mixing angles $(\sin^2\theta_{12},\,\sin^2\theta_{23},\,\sin^2\theta_{13})$,
as well as on three observables sensitive to absolute $\nu$ masses: the effective mass
$m_\beta$ in single beta  decay, the effective Majorana mass $m_{\beta\beta}$ 
in neutrinoless double beta ($0\nu2\beta$) decay, and the sum of $\nu$ masses $\Sigma$ in
cosmology---see  \cite{Fogli2006,Fogli2004,Foglireview} for notation and conventions. We update
the results of \cite{Fogli2006} by including several new experimental inputs, largely 
presented or discussed 
at the recent {\em Neutrino 2008\/} Conference
\cite{Nu2008}.

\medskip\medskip
{\em Neutrino oscillation updates.}
The Kamioka Liquid Scintillator Anti-Neutrino
Detector (KamLAND) Collaboration has presented reactor $\overline\nu_e$ disappearance
and geo-$\nu$ results for an exposure of 2.881 kTy \cite{KL2008}, a factor $\sim\!\!4$ higher
than the one we used in \cite{Fogli2006}. 
Following \cite{KL2008}, the KamLAND spectrum
analysis in \cite{Fogli2006,Foglireview} has been upgraded \cite{Rotunno} 
to include the rates of geo-$\nu$ events from U and Th decay as low-energy
nuisance parameters.

Results from the third phase of the Sudbury Neutrino Observatory (SNO-III)
\cite{SNO3}, recently presented at {\em Neutrino 2008} \cite{Nu2008}, have been
included \cite{Rotunno} in the form of two new integral determinations of the 
charged- and neutral-current event rates \cite{SNO3}.  
Other solar $\nu$ updates, with a minor impact in the global parameter 
estimate, include the latest Borexino results \cite{Borexino} and reevaluated 
GALLEX data \cite{GALLEX}---see also \cite{theta13}.

The Main Injector Neutrino Oscillation Search
(MINOS) Collaboration has presented accelerator $\nu_\mu$ disappearance data 
from $3.36\times 10^{20}$ protons on target \cite{MINOS2008}, a factor of $\sim$2.6 larger
than previously used in \cite{Fogli2006}. In the official MINOS data
analysis \cite{MINOS2008}, for any given energy profile of the $\nu_\mu$ survival probability $P_{\mu\mu}(E_\nu)$,
a ``beam matrix'' method is used to map the energy spectrum from near to far, and 
an independent near-far extrapolation method is used as a cross-check
\cite{MINOSlast}. This approach can be fully implemented only within the Collaboration.
For our purposes, we analyze the 
18-bin energy spectrum ratio \cite{MINOS2008} by folding  
the function $P_{\mu\mu}(E_\nu,\Delta m^2,\,\sin^2\theta_{23},\,\sin^2\theta_{13})$, 
with empirical energy resolution profiles, which mimic the near-far energy spectrum mapping 
of \cite{MINOSlast}. Normalization and energy scale systematics are treated as nuisance parameters. 

In the limit $\theta_{13}\to 0$,
our effective $2\nu$ parameter fits reproduce very well the official ones 
as obtained by the KamLAND \cite{KL2008},
SNO-III \cite{SNO3}, and MINOS \cite{MINOS2008} Collaborations. In our
global analysis, however, we treat $\theta_{13}$ as a free parameter.

\begin{figure}
\vspace*{4cm}
\includegraphics[width=\textwidth]{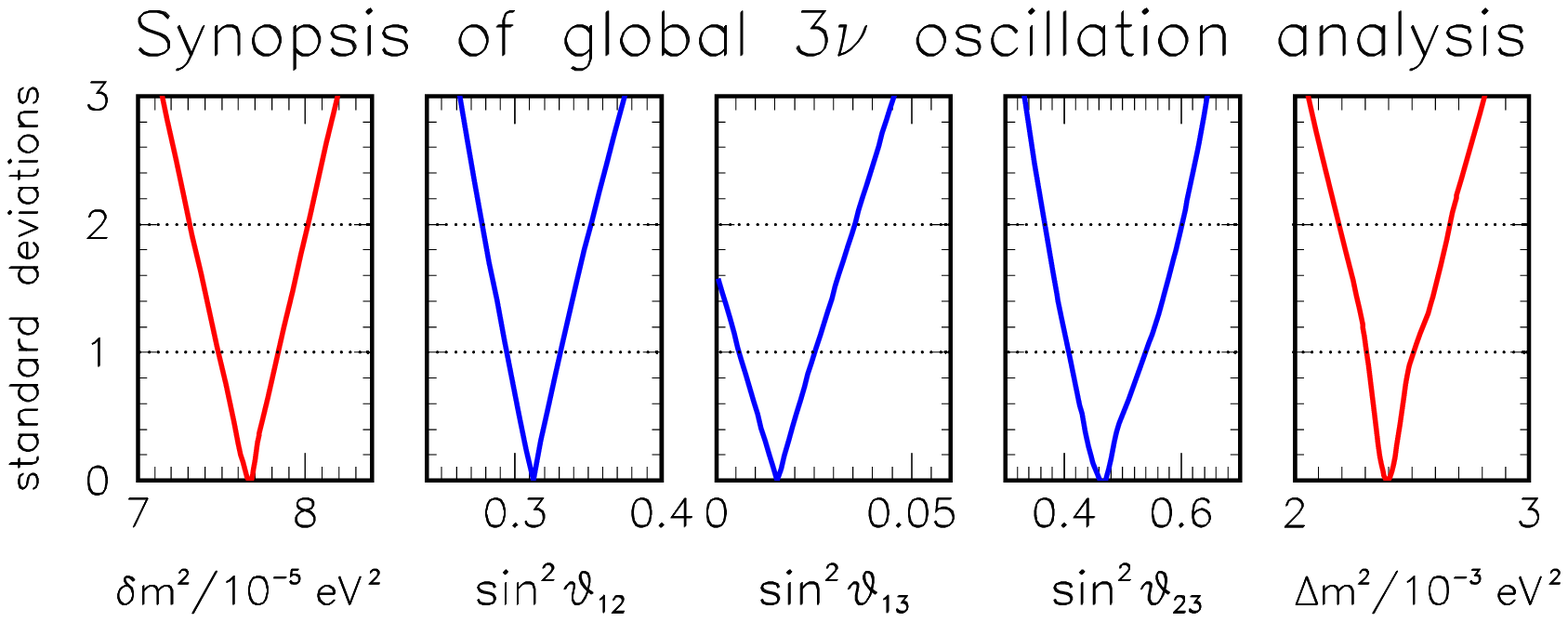}
\vspace*{0.5cm}
\caption{\label{fig_01} 
Global $3\nu$ oscillation analysis (2008): Bounds on the 
mass-mixing oscillation parameters, in terms of standard deviations  from 
the best fit. Note the $1.6\sigma$ preference for $\theta_{13}>0$.} 
\end{figure}

Figure~\ref{fig_01} displays our updated results on the mass-mixing parameters, in terms of
standard deviations $n_\sigma$ from the best fit ($n_\sigma=\sqrt{\Delta\chi^2}$ after $\chi^2$ marginalization).
Table~\ref{Synopsis} summarizes such results in numerical form.
As compared with \cite{Fogli2006}, the $\Delta m^2$ uncertainty
is almost
halved (by new MINOS data), and both the $\delta m^2$ and the $\sin^2 2\theta_{12}$ allowed
ranges are reduced (by new KamLAND and SNO data). The range of $\sin^2\theta_{23}$ is almost unchanged.
As discussed in \cite{theta13}, an intriguing new result
is the preference for $\theta_{13}>0$ at the level of
$\sim\!\!1.6\sigma$ (or, equivalently,  $\sim\! 90\%$~C.L.). Such an indication
emerges from the combination of two independent hints in favor of $\theta_{13}>0$, each 
at the level of $\sim\!\!1\sigma$: an older one, found in the atmospheric $\nu$
data analysis of \cite{Foglireview}, and a newer one, coming from the
small difference between the best-fit values of 
$\sin^22\theta_{12}$ in KamLAND \cite{KL2008} and SNO \cite{SNO3}---a difference which
is reduced for $\sin^2\theta_{13}\sim \mathrm{few}\ \%$.
Hereafter, as in  \cite{Fogli2006,Fogli2004},
we shall show results at a conservative $2\sigma$ (95\%) confidence
level, in which case only an upper bound can be placed on $\theta_{13}$.

\begin{figure}
\vspace*{2.5cm}
\includegraphics[width=0.9\textwidth]{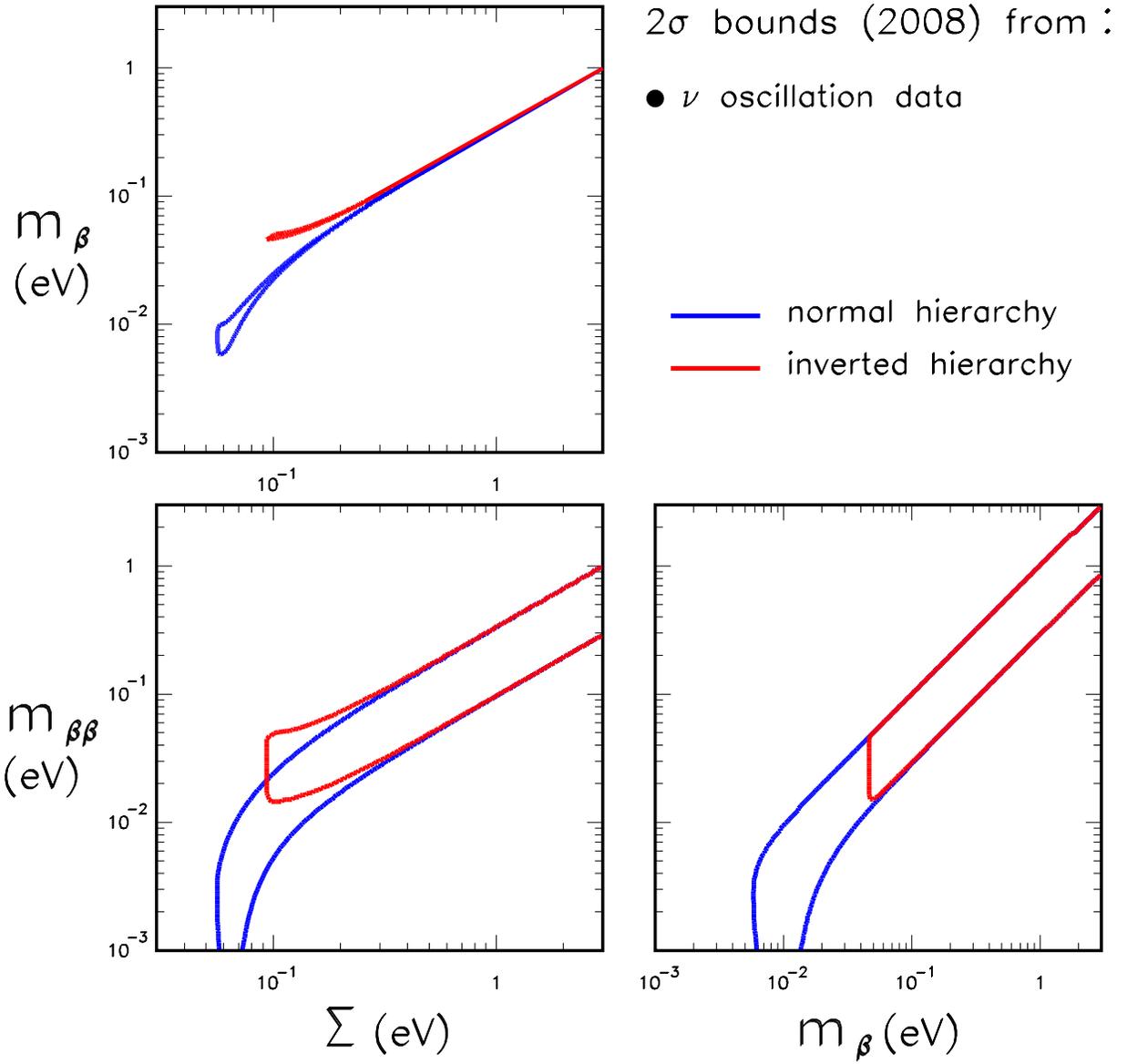}
\vspace*{0.5cm}
\caption{\label{fig_02} 
Bands allowed at $2\sigma$ by neutrino oscillation data, in each of the
three coordinate planes of the parameter  space ($m_\beta,\,m_{\beta\beta},\,\Sigma$),
for both normal and inverted hierarchy.}
\end{figure}

Figure~\ref{fig_02} shows the
$2\sigma$ bounds implied by the above $\nu$ oscillation parameter constraints
(for normal or inverted hierarchy)
in the three planes charted by any two among the three observables
$(m_\beta,\,m_{\beta\beta},\Sigma)$. A measurement of any such quantity, coupled with the
bounds in Fig.~\ref{fig_02}, provides ``predictions'' for the other two quantities \cite{Fogli2006,Fogli2004,Foglireview}.

\medskip\medskip

{\em Cosmology updates.}
Within the
standard cosmological model,
the 5-year data recently released  by Wilkinson Microwave Anisotropy Probe (WMAP~5y)  \cite{WMAP1,WMAP2} constrain, by themselves, the
sum of the $\nu$ masses $\Sigma$ below 1.3~eV  at 95\% C.L.\ \cite{WMAP1}. This limit can be
strengthened in the sub-eV range by adding further cosmological data; for instance, the
WMAP collaboration finds $\Sigma<0.61$~eV 
by adding Baryonic Acoustic Oscillation (BAO) and type-Ia supernova (SN-Ia) data \cite{WMAP2}.

We consider five representative combinations of cosmological data, which lead to 
increasingly stronger upper limits on $\Sigma$: 
(1) Cosmic Microwave Background (CMB) anisotropy data from: WMAP~5y
\cite{WMAP2}, Arcminute Cosmology Bolometer Array Receiver (ACBAR) 
\cite{acbar07}, Very Small Array (VSA) \cite {vsa}, 
  Cosmic Background Imager (CBI) \cite{cbi} and
  BOOMERANG \cite{boom03} experiments;
(2) the above CMB results plus the large-scale structure
(LSS) information on galaxy clustering coming from the Luminous Red Galaxies Sloan Digital Sky Survey 
(SDSS) \cite{Tegmark}; 
(3) the above CMB results plus the Hubble Space Telescope (HST) prior on the value of the reduced
  Hubble constant, $h=0.72 \pm0.07$ \cite{hst}, and the luminosity
  distance SN-Ia data of \cite{astier};
(4) the data in (3) plus the BAO data  from \cite{bao};
(5) the data in (4), plus the small scale primordial spectrum from 
Lyman-alpha (Ly$\alpha$) forest clouds \cite{Ly1,Ly2}.

\begin{figure}
\vspace*{2cm}
\includegraphics[width=0.9\textwidth]{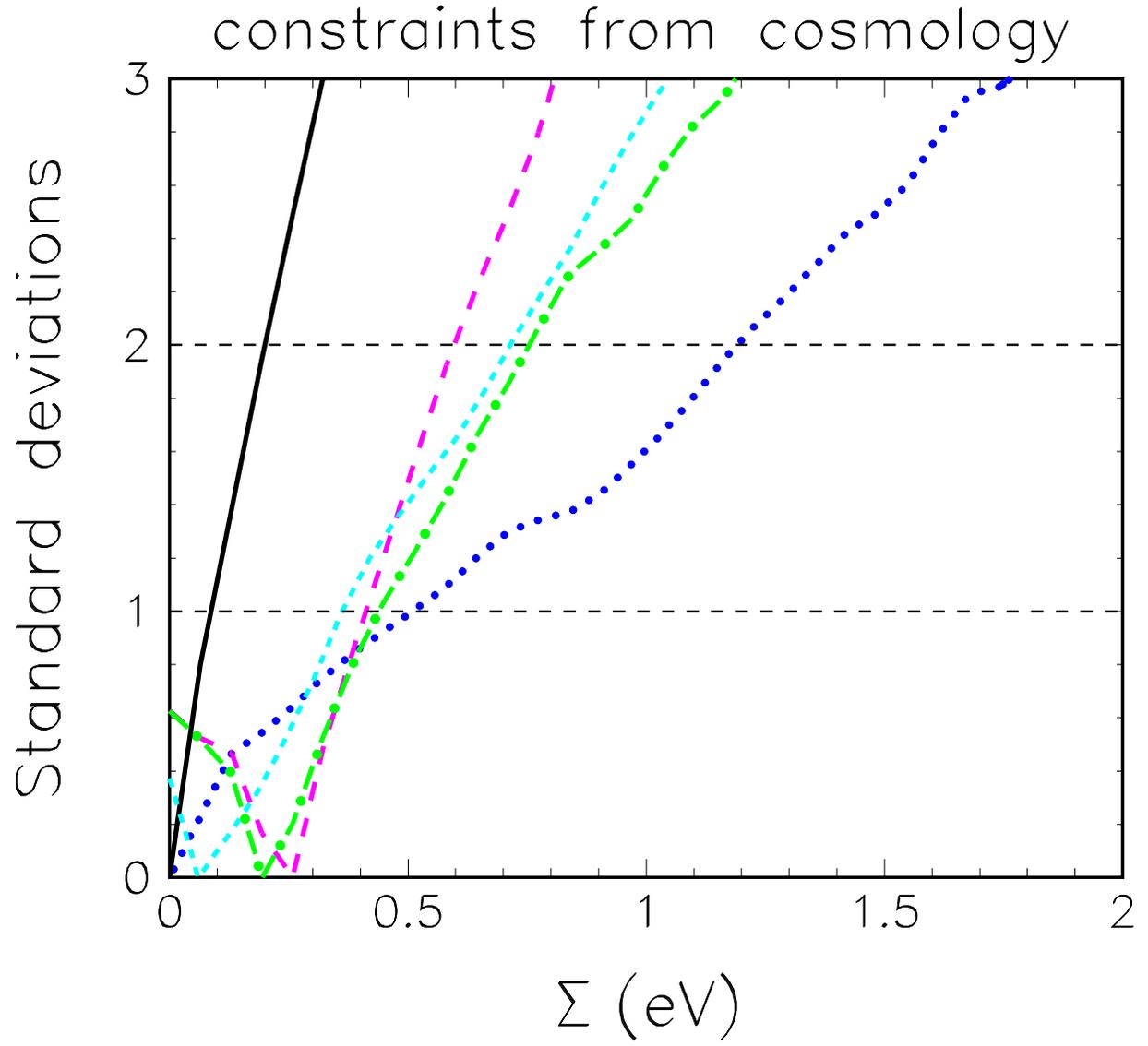}
\vspace*{0.5cm}
\caption{\label{fig_03} 
Cosmological constraints on the sum of neutrino masses ($\Sigma$). 
Standard deviation curves for the five datasets in Table~\protect\ref{tableCASES}:
1 (dotted), 2 (dashed), 3 (dot-dashed), 4 (long dashed), and 5 (solid).}
\end{figure}

Adopting the same procedure described in \cite{Fogli2006}, based on the
publicly available COSMOMC code \cite{lewis}, we find 
the upper limits on $\Sigma$ summarized in Table~\ref{tableCASES} and in Fig.~\ref{fig_03}. 
The bound in Table~\ref{tableCASES} for case (1) is dominated by WMAP-5y data. The results for cases (1)
and (4) are in agreement with similar constraints presented
in \cite{WMAP1} and \cite{WMAP2}, respectively, even if the datasets considered here
for BAO and SN-Ia are different. In Fig.~\ref{fig_03}, the slight preference for $\Sigma\neq 0$ at best
fit for case (4) [and case (1)], also found in \cite{WMAP2}, is not statistically
significant. 
As in Ref.~\cite{Fogli2006}, we find that Ly$\alpha$ data have a strong impact on $\Sigma$,
but their inclusion in global fits is debated \cite{WMAP2} due to systematics still under
scrutiny. The upper limits from cases (1)--(4) (namely, $\Sigma<0.6-1.2$~eV) should be considered as
more conservative. Including LSS data 
would not significantly modify case~5, which is dominated by Ly$\alpha$ data. 
In the following, we shall focus on the two extreme cases (1) and (5).

\begin{table}[t]
\caption{\label{Synopsis} Global $3\nu$ oscillation analysis (2008): best-fit values and
allowed $n_\sigma$ ranges  for the mass-mixing parameters.}
\resizebox{\textwidth}{!}{
\begin{ruledtabular}
\begin{tabular}{cccccc}
Parameter & $\delta m^2/10^{-5}\mathrm{\ eV}^2$ & $\sin^2\theta_{12}$ & $\sin^2\theta_{13}$ & $\sin^2\theta_{23}$ &
$\Delta m^2/10^{-3}\mathrm{\ eV}^2$ \\[4pt]
\hline
Best fit        &     7.67     &  0.312          &  0.016          &  0.466          &  2.39 \\
$1\sigma$ range & 7.48~--~7.83 & 0.294~--~0.331  & 0.006~--~0.026  & 0.408~--~0.539  & 2.31~--~2.50 \\
$2\sigma$ range & 7.31~--~8.01 & 0.278~--~0.352  & $<0.036$        & 0.366~--~0.602  & 2.19~--~2.66 \\
$3\sigma$ range & 7.14~--~8.19 & 0.263~--~0.375  & $<0.046$        & 0.331~--~0.644  & 2.06~--~2.81 
\end{tabular}
\end{ruledtabular}
}
\end{table}

\begin{table}[b]
\caption{\label{tableCASES} Representative cosmological data sets 
and corresponding $2\sigma$ (95\% C.L.) constraints on the sum of $\nu$ masses $\Sigma$.}
\resizebox{\textwidth}{!}{
\begin{ruledtabular}
\begin{tabular}{llr}
Case 	& Cosmological data set		& $\Sigma$ (at $2\sigma$)\\[4pt]
\hline
1 		& CMB  													& $<1.19$ eV \\
2		 & CMB + LSS						                        & $<0.71$ eV \\
3		& CMB + HST + SN-Ia					                    & $<0.75$ eV \\ 
4 		& CMB + HST + SN-Ia + BAO			                    & $<0.60$ eV \\ 
5		& CMB + HST + SN-Ia + BAO + Ly$\alpha$                 & $<0.19$ eV
\end{tabular}
\end{ruledtabular}
}
\end{table}

\medskip\medskip

{\em $0\nu2\beta$ decay updates.}
The final analysis of part of the Heidelberg-Moscow (HM) Collaboration reports a $0\nu2\beta$ signal in $^{76}$Ge
with half-life $T^{0\nu}_{1/2}=2.23^{+0.44}_{-0.31}\times 10^{25}$~y ($1\sigma$ errors) at a claimed 
C.L.\ $>6\sigma$ \cite{Klapdor2006}. 
The previously estimated $T^{0\nu}_{1/2}$ \cite{Klapdor2004}, as used in \cite{Fogli2006}, was
a factor of $\sim$2 smaller. The claim is controversial, but the experimental sensitivity to the signal 
(if real) is no longer questioned \cite{Cuoricino}.

From a theoretical viewpoint, the $0\nu2\beta$ nuclear matrix elements (NME) $C_{mm}$ and uncertainties
estimated via Quasiparticle Random Phase Approximations (QRPA)
in \cite{Rodin2007} (as used in \cite{Fogli2006})  have been  recently revised 
\cite{RodinErr,Simkovic2007}, especially to improve the so-called short-range correlations. 
We adopt for $C_{mm}$ 
the central values and errors of \cite{Simkovic2007}, which agree
with independent QRPA \cite{Suhonen1,Suhonen2} and shell model \cite{Poves} evaluations within 
$\sim$2$\sigma$ (see Fig.~11 of \cite{Simkovic2007} and related comments therein).

The effect of both the $T^{0\nu}_{1/2}$ and the NME 
updates for $^{76}$Ge is to lower the central value---and to enlarge
the errors---of the effective mass parameter $m^2_{\beta\beta}=m^2_e/C_{mm}T^{0\nu}_{1/2}$. 
By taking logs (in base 10) to linearize error
propagation, 
 we have $\log (T^{0\nu}_{1/2}/\mathrm{y})=23.35\pm 0.16$ ($2\sigma$) from 
\cite{Klapdor2006} and $\log(C_{mm}/\mathrm{y}^{-1})=-12.82\pm 0.48$ 
($2\sigma$) from \cite{Simkovic2007}, so that
\begin{equation}
\log (m_{\beta\beta}/\mathrm{eV}) = -0.54 \pm 0.26 \ \ \ (\mathrm{HM\ claim},\; 2\sigma)\ ,
\label{HMclaim}
\end{equation}
where the experimental error and the (dominant) theoretical error have been added in quadrature.

The Cuoricino experiment, which does not find $0\nu2\beta$ decay signals  in $^{130}$Te,   
quotes $T^{0\nu}_{1/2}>3.1\times 10^{24}$~y at
90\% C.L.\ \cite{Cuoricino}, or $T^{0\nu}_{1/2}>2.5\times 10^{24}$~y at 95\% C.L.\ \cite{Cremonesi}. Using  the
latter limit as $\log(T^{0\nu}_{1/2}/\mathrm{y})>24.4$,
and the $^{130}$Te  NME estimate $\log(C_{mm}/\mathrm{y}^{-1})=-12.27\pm 0.28\; (2\sigma)$
from \cite{Simkovic2007}, we get 
\begin{equation}
\log (m_{\beta\beta}/\mathrm{eV}) < [-0.63,\,-0.07] \ \ \ (\mathrm{Cuoricino},\; 2\sigma)\ ,
\end{equation}
where the range due to the $2\sigma$ uncertainty of the NME is explicitly reported.

A comparison of the corresponding $m_{\beta\beta}$ ranges ($2\sigma$),
\begin{eqnarray}
0.16 < m_{\beta\beta}/\mathrm{eV} < 0.52 &&  \ \ \ (\mathrm{HM\ claim})\ ,\label{HMclaim2}\\
0\leq  m_{\beta\beta}/\mathrm{eV} < 0.23 && \ \ \ (\mathrm{Cuoricino,\ ``favorable"\ NME})\ ,\\
0\leq  m_{\beta\beta}/\mathrm{eV} < 0.85 && \ \ \ (\mathrm{Cuoricino,\ ``unfavorable"\ NME})\ ,
\end{eqnarray}
shows that current Cuoricino data may or may not disfavor a fraction of the HM 
range for $m_{\beta\beta}$ at $2\sigma$, depending on the (still quite uncertain)
value of the $^{130}$Te $0\nu2\beta$ NME.   
 A similar conclusion (albeit with somewhat
different preferred ranges for $m_{\beta\beta}$) has been reached in \cite{Cuoricino}. 
 Therefore, the $0\nu2\beta$ claim \cite{Klapdor2006}
remains an open issue at present, and we shall consider the possibility that
it corresponds to a real signal.

\begin{figure}
\vspace*{2cm}
\includegraphics[width=0.9\textwidth]{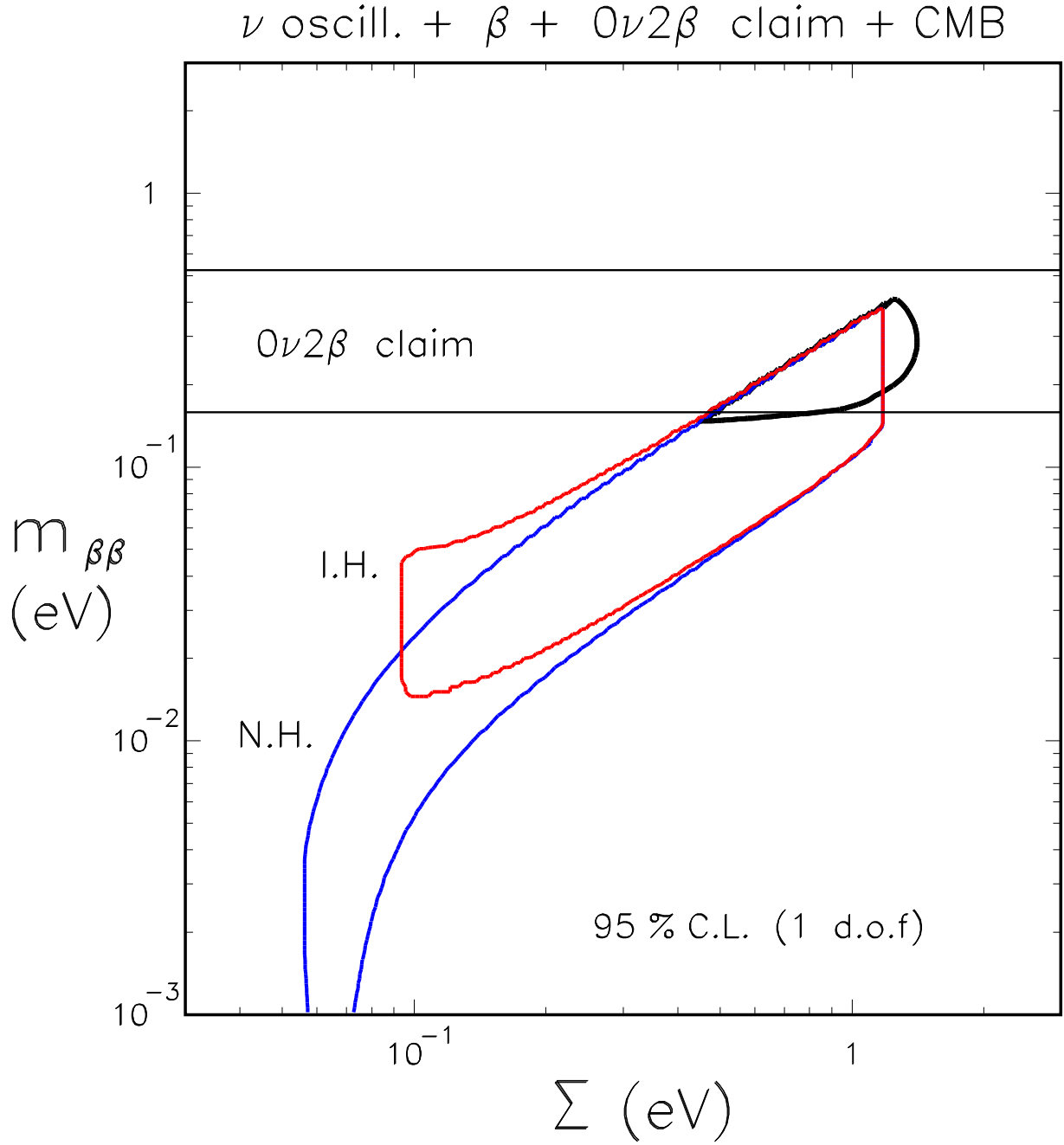}
\vspace*{0.5cm}
\caption{\label{fig_04} 
Global combination of oscillation plus CMB data (case 1 in Table~\protect\ref{tableCASES}) with the
$0\nu2\beta$ decay claim, in the plane $(\Sigma,\,m_{\beta\beta})$.}
\end{figure}

\medskip\medskip
{\em Discussion.}
Figure~\ref{fig_04} shows the regions allowed at $2\sigma$ in normal and inverted hierarchy (slanted bands)
by the combination of oscillation results with the first dataset in Table~\ref{tableCASES} (CMB), 
in the plane spanned by ($\Sigma,\,m_{\beta\beta}$). This is the most conservative
case, with the weakest limits on $\Sigma$, and the largest
overlap between the regions separately allowed by oscillation+CMB data and
by the $0\nu2\beta$ claim. The results of a global $\chi^2$ fit are shown as a thick black
wedge in the upper right part of the figure.
[The combination includes the current limit $m_{\beta}<1.8$~eV ($2\sigma$) 
\protect\cite{Fogli2006} which, however, provides only a minor contribution.]
 Such global combination would correspond  to nearly degenerate masses in the range
$$
m_1\simeq m_2\simeq m_3 \in [0.15,\,0.46]~\mathrm{eV}\ \ (2\sigma)\ . 
$$

In this case (degenerate spectrum), the preferred range for effective neutrino mass in $\beta$ decay would also
be $m_\beta\in [0.15,\,0.46]$~eV.
In the upper half of this range, the KArlsruhe TRItium Neutrino (KATRIN) $\beta^-$ experiment could make a
$5\sigma$ discovery, according to the estimated sensitivity \cite{KATRIN}. A $3\sigma$ evidence
could still be found in KATRIN for $m_\beta\sim 0.3$~eV. Below this value, the sensitivity would
be rapidly degraded, and only upper bounds could be placed for $m_\beta\lesssim 0.2$~eV \cite{KATRIN}.
The possibility of reaching a $\sim$0.1--0.2~eV sensitivity with a different approach to
$\beta$ decay is being
discussed \cite{MARE}.

If the cosmological dataset (1) were replaced by the datasets (2)--(4) in Table~\ref{tableCASES}, the
overlap region between  the $0\nu2\beta$ band and the oscillation+cosmological bands in Fig.~\ref{fig_04}
would shrink (not shown), but would not disappear. 
Therefore, within the standard $3\nu$ framework and the present uncertainties, 
the $0\nu2\beta$ claim clashes with
oscillation+cosmological data only if the latter include Ly$\alpha$ data.

\begin{figure}
\vspace*{2cm}
\includegraphics[width=0.9\textwidth]{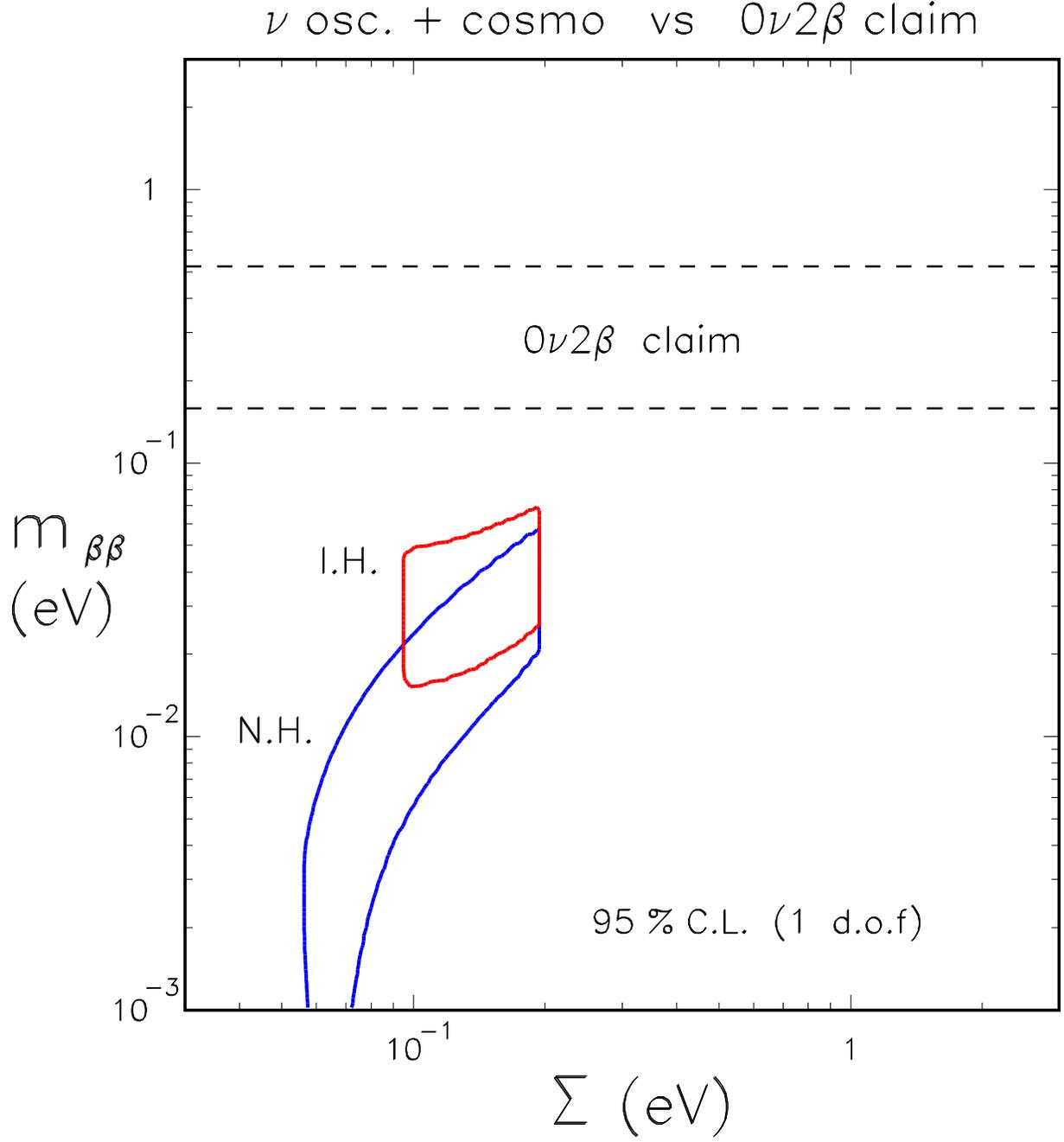}
\vspace*{0.5cm}
\caption{\label{fig_05} 
Bounds from oscillation plus all cosmological data (case 5 in Table~\protect\ref{tableCASES}), contrasted with the
$0\nu2\beta$ decay claim, in the plane $(\Sigma,\,m_{\beta\beta})$.}
\end{figure}

Figure~\ref{fig_05} is analogous to Fig.~\ref{fig_04}, but refers to the fifth dataset in Table~\ref{tableCASES} (all cosmological
data, including Ly$\alpha$). In this case, the allowed regions do not overlap and cannot be combined, 
since the relatively strong cosmological limit $\Sigma<0.19$~eV
implies $m_{\beta\beta}\lesssim 0.08$~eV, in contradiction with Eq.~(\ref{HMclaim2}).
Solutions to this discrepancy would require that either some data or their interpretation are wrong.

In conclusion, important pieces of information are being slowly added to the puzzle of 
absolute $\nu$ masses. In this followup to \cite{Fogli2006}, we have discussed the most recent 
oscillation and nonoscillation updates in the field, after the recent  {\em Neutrino 2008\/} Conference
\cite{Nu2008}. Oscillation parameters are robustly constrained,
and an intriguing indication for  $\theta_{13}>0$ emerges, as summarized in 
Fig.~\ref{fig_01} and in Table~\ref{Synopsis}. 
Concerning nonoscillation observables, despite some recent
experimental and theoretical progress, a coherent picture remains 
elusive. In particular, the $0\nu2\beta$ claim is still under independent experimental scrutiny, and
it may be compatible (Fig.~\ref{fig_04}) or incompatible (Fig.~\ref{fig_05}) with the 
cosmological bounds (Table~\ref{tableCASES}), depending on data selection (especially Ly$\alpha$).
A confident assessment of the $\nu$ mass
scale will require converging evidence from at least two 
of the three observables $(m_\beta,\,m_{\beta\beta},\,\Sigma)$ within the
bands allowed by oscillation data in Fig.~\ref{fig_02}.

\newpage
\medskip\medskip
{\em Acknowledgments.}
G.L.\ Fogli, E.\ Lisi, A.\ Marrone, and A.\ Rotunno acknowledge 
support by the Italian MIUR and INFN through the ``Astroparticle Physics'' 
research project, and by the EU ILIAS through the ENTApP project. 
A.~Palazzo is supported by MEC under the I3P program, by Spanish grants FPA2005-01269
and by European Commission network MRTN-CT-2004-503369 and
ILIAS/N6 RII3-CT-2004-506222. This research has also
been supported by ASI contract I/016/07/0 ``COFIS.''

\newpage
\newpage
\newpage
\newpage
\newpage

\end{document}